\documentclass[prl,superscriptaddress,amsmath,amssymb,floatfix,twocolumn,amsfonts]{revtex4-2}

\usepackage{times}
\usepackage[varg]{txfonts}
\usepackage{textcomp}
\usepackage{graphicx}
\usepackage{subfigure}
\usepackage{subeqnarray}
\usepackage{mathtools}
\usepackage{color}
\usepackage[colorlinks=true,citecolor=blue,urlcolor=blue,linkcolor=blue,hyperindex]{hyperref}
\usepackage{braket}
\usepackage{overpic}
\usepackage{ulem}
\usepackage[version=3]{mhchem}

\newcommand{\al}{{\alpha}}

\newcommand{\bk}{{\bf k}}

\def\be{\begin{equation}}       \def\ee{\end{equation}}
\def\bea{\begin{eqnarray}}      \def\eea{\end{eqnarray}}

\begin{document}

\title{Superconducting Klein and anti-Klein tunneling in Weyl junctions}

\author{Jiajia Huang}
\affiliation{State Key Laboratory of Optoelectronic Materials and Technologies, Guangdong Provincial Key Laboratory of Magnetoelectric Physics and Devices, School of Physics, Sun Yat-Sen University, 510275 Guangzhou, China}
\author{Luyang Wang}
\email{wangly@szu.edu.cn}
\affiliation{College of Physics and Optoelectronic Engineering, Shenzhen University, Shenzhen 518060, China}
\author{Dao-Xin Yao}
\email{yaodaox@mail.sysu.edu.cn}
\affiliation{State Key Laboratory of Optoelectronic Materials and Technologies, Guangdong Provincial Key Laboratory of Magnetoelectric Physics and Devices, School of Physics, Sun Yat-Sen University, 510275 Guangzhou, China}
%\affiliation{State Key Laboratory of Optoelectronic Materials and Technologies, School of Physics, Sun Yat-sen University, Guangzhou 510275, China}
%\affiliation{International Quantum Academy, Shenzhen 518048, China}

\date{\today}

\begin{abstract}
Klein tunneling is an old topic in relativistic quantum physics, and has been observed recently in graphene where massless particles reside. Here, we propose a new heterostructure platform for Klein tunneling to occur, which consists of a Weyl-semimetal-based normal state/superconductor (NS) junction. By developing a Blonder-Tinkham-Klapwijk-like theory, we find that Klein tunneling occurs at normal incidence, which can lead to differential conductance doubling. If the (single) Weyl semimeltals are replaced by double Weyl semimetals, anti-Klein tunneling will take place of Klein tunneling. Our work provides a theoretical guide for the detection of (anti-)Klein tunneling in three-dimensional chiral NS junctions.
\end{abstract}

\maketitle

%%%%%%%%%%%%%%%%%%%%%%%%%%%%%%%%%%%%%%%%%%%%%%%%%%%%%%%%%%%%%%%
\textit{Introduction.}---
Klein tunneling is a fascinating quantum mechanical phenomenon that occurs in relativistic systems, where particles can perfectly tunnel through potential barriers with vanishing reflectivity\cite{Klein,Rose1961,Holstein1998,Dombey1999,Calogeracos1999HistoryAP}. It leads to many interesting phenomena in nuclear and particle physics\cite{Reinhardt1977}, and has been used to explain the evaporation of black holes\cite{Parikh2000,Page2005}. In recent years, as massless relativistic particles are found in condensed matter systems\cite{Vafek}, Klein tunneling has been predicted to exist in graphene\cite{NatureKatsnelson,Allain2011} and Weyl semimetals (WSM)\cite{Spivak2016,Beenakker2016,Yesilyurt2016,Bai2016,Nguyen2018}, and observed indirectly\cite{Stander2009,Andrea2009,Beenakker,Young2011} or directly\cite{Elahi2024} in graphene-based systems. It allows for the creation of electron waveguides, where electrons can propagate along specific paths without being affected by potential barriers\cite{cheianov2007,Wilmart2014}. As such, Klein tunneling plays a crucial role in the transport properties and electronic conductivity of graphene, making it a promising material for various applications in nanoelectronics.

Klein tunneling has also been shown to exist in normal state/superconducting (NS) junctions\cite{SLee2019}. The normal (N) side consists of topological surface states (TSS) while on the superconductor (S) side lie TSS in proximity to an s-wave superconductor. Perfect Andreev reflection, where incident electrons are totally reflected as holes, occurs at normal incidence, due to spin-momentum locking which leads to perfect matching of incident electron and reflected hole wavefunctions at the interface.

In this paper, we dig deeper into Klein tunneling in NS junctions. In particular, we investigate the tunneling in WSM-based NS junctions. By developing Blonder-Tinkham-Klapwijk (BTK)-like theory\cite{BTK} in this system, we find that Klein tunneling indeed occurs at normal incidence from a WSM to an s-wave superconductivity proximitized WSM (sWSM), which can give rise to differential conductance doubling. We also consider replacing the WSM by a double WSM (DWSM) and the sWSM by an s-wave superconductivity proximitized DWSM (sDWSM). Interestingly, anti-Klein tunneling occurs in this DWSM/sDWSM system -- electrons are completely reflected as the delta function potential barrier at the interface becomes strong. %Similar behavior has been predicted and observed in graphitic systems\cite{NatureKatsnelson,}., and here we predict it in a new platform. It has important implications for our understanding of relativistic systems and has paved the way for numerous technological advancements in the field of two-dimensional materials.

\begin{figure}
    \centering
    \includegraphics[width=0.98\linewidth]{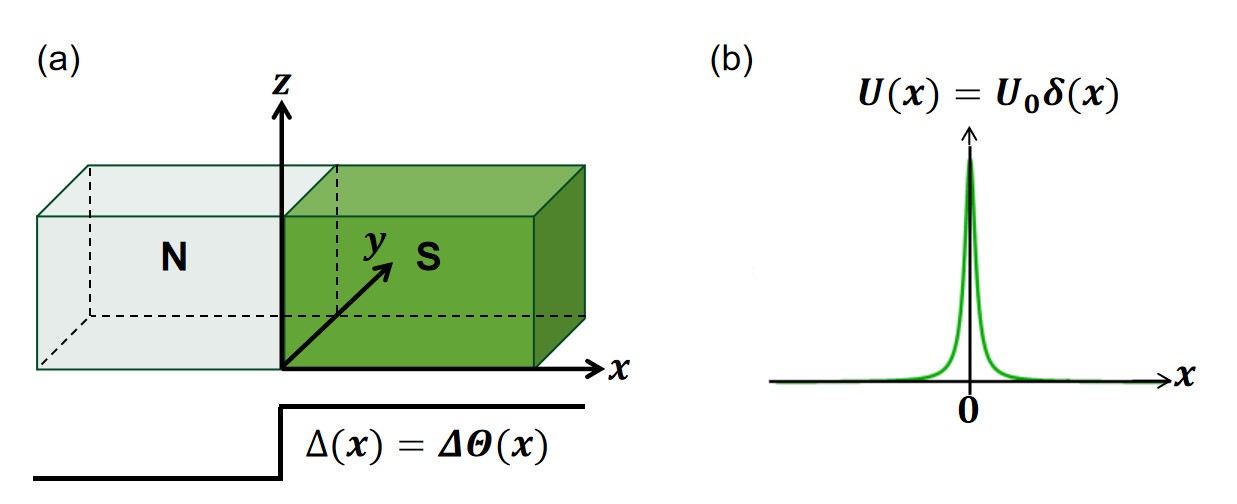}
    \caption{(a) The NS junction under study. On the N (S) side is a WSM (sWSM). (b) The interface at $x=0$ is modeled by a delta function potential barrier, in the spirit of the BTK model.}
    \label{fig:junction}
\end{figure}

\begin{figure}
    \centering
    \includegraphics[width=0.98\linewidth]{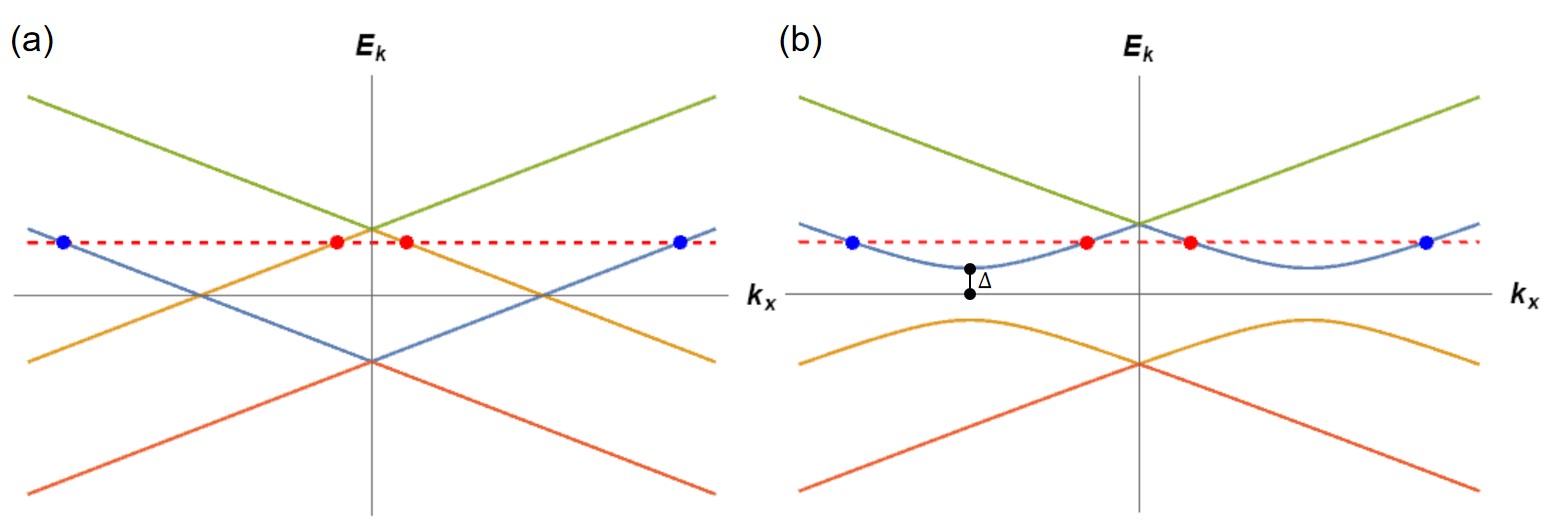}
    \caption{(a) Dispersion of a (single) WSM along the $k_x$-axis in Nambu space. (b) Dispersion of an sWSM along the $k_x$-axis. The blue (red) dots indicate the eletron-like (hole-like) propagating wavefunctions at $\Delta<E<\mu$.}
    \label{fig:single}
\end{figure}

\textit{Klein tunneling in WSM-sWSM junctions.}---Our model consists of a WSM on the N side, and an sWSM on the S side, as shown in Fig.\ref{fig:junction}(a). In the spirit of the BTK model, the interface is modeled by a delta function potential barrier, as shown in Fig.\ref{fig:junction}(b). Therefore, the Hamiltonian of the heterostructure reads
\begin{eqnarray}
H_{hetero}=\left(
    \begin{array}{cc}
      v_{F}\bk\cdot\boldsymbol{\sigma}-\sigma_{0}\mu+\sigma_{0}U(x) & i\sigma_{y}\Delta(x) \\
      -i\sigma_{y}\Delta(x) & v_{F}\bk\cdot\boldsymbol{\sigma}^{*}+\sigma_{0}\mu-\sigma_{0}U(x) \\
    \end{array}
  \right),\nonumber\label{eq:H_hetero}\\
\end{eqnarray}
where $v_{F}\bk\cdot\boldsymbol{\sigma}$ describes Weyl fermions with Pauli matrices representing (pseudo)spin, $\mu$ is the chemical potential, $U(x)=U_0\delta(x)$, and $\Delta(x)=\Delta\Theta(x)$ where $\Theta(x)$ is the Heaviside step function. The dispersions on the two sides are shown in Fig.\ref{fig:single}. On the N side, the eigenenergies of the electron-like states are $E_{k,\pm}-\mu=\pm v_{F}k-\mu$, with corresponding eigenstates
\bea
\Psi_{e,\pm}&=&\frac{1}{\sqrt{1+\eta_{\pm}^2}}\left(
                             \begin{array}{c}
                               1 \\
                               \eta_\pm e^{i\theta_{k}} \\
                             \end{array}
                           \right),
\eea
where $\eta_\pm=\pm\sqrt{\frac{E_{k,\pm}-v_{F}k_z}{E_{k,\pm}+v_{F}k_z}}=\pm\sqrt{\frac{\pm1-\cos\alpha_k}{\pm1+\cos\alpha_k}}$
with $\alpha_k=\cos^{-1}(k_z/k)$, and $\theta_k=\tan^{-1}(k_y/k_x)$. Then $\eta_+=\tan\frac{\alpha_k}{2}$ and $\eta_-=-\cot\frac{\alpha_k}{2}$. We will consider an incident electron in the upper band with energy $E$, whose wavefunction is
\begin{eqnarray}
  \psi_{in} &=& \cos\frac{\alpha_k}{2}\left(
                             \begin{array}{c}
                               1 \\
                               \tan\frac{\alpha_k}{2} e^{i\theta_{k}} \\
                             \end{array}
                           \right)
           =\left(
                             \begin{array}{c}
                               \cos\frac{\alpha_k}{2} \\
                               \sin\frac{\alpha_k}{2} e^{i\theta_{k}} \\
                             \end{array}
                           \right).
\end{eqnarray}
The electron can be reflected either to an electron state or a hole state. For a reflected electron, we need to replace $k_{x}$ by $-k_{x}$ in the wavefunction, which is equivalent to replace $e^{i\theta_{k}}$ by $-e^{-i\theta_{k}}$, yielding the reflected wavefunction
\begin{eqnarray}
 \left(
 \begin{array}{c}
 \cos\frac{\alpha_k}{2} \\
  -\sin\frac{\alpha_k}{2} e^{-i\theta_{k}} \\
   \end{array}
   \right).
\end{eqnarray}
For a reflected hole, the momentum is the same as the incident electron while the energy is opposite. Then the wavefunction is
\begin{eqnarray}
  \left(
 \begin{array}{c}
 -\sin\frac{\alpha_k}{2}e^{i\theta_k} \\
  \cos\frac{\alpha_k}{2}  \\
   \end{array}
   \right).
\end{eqnarray}
Therefore, on the N side, the whole wavefunction consists of three parts,
\begin{eqnarray}
\Psi_{N}&=&\left(\begin{array}{c}
\cos\frac{\alpha_k}{2} \\
\sin\frac{\alpha_k}{2} e^{i\theta_{k}} \\
0\\
0
\end{array}\right)
+r_{e}\left(
 \begin{array}{c}
 \cos\frac{\alpha_k}{2} \\
  -\sin\frac{\alpha_k}{2} e^{-i\theta_{k}} \\
  0\\
  0
   \end{array}
   \right)
+  r_h\left(
 \begin{array}{c}
 0\\
 0\\
 -\sin\frac{\alpha_k}{2}e^{i\theta_k} \\
  \cos\frac{\alpha_k}{2}
   \end{array}
   \right),\nonumber\\
\end{eqnarray}
where $r_e$ and $r_h$ are the reflection coefficients of the corresponding reflecting processes. On the S side, the wavefunction corresponding to the same energy $E$ with the assumption $E>\Delta$ consists of two parts, 
\begin{align}
    \Psi_S=t_e\begin{pmatrix}
       u\cos{\frac{\al_k}{2}}\\
       u\sin{\frac{\al_k}{2}}e^{i\theta_k}\\
       -v\sin{\frac{\al_k}{2}}e^{i\theta_k}\\
       v\cos{\frac{\al_k}{2}}
    \end{pmatrix}+t_h\begin{pmatrix}
        v\cos{\frac{\al_k}{2}}\\
        -v\sin{\frac{\al_k}{2}}e^{-i\theta_k}\\
        u\sin{\frac{\al_k}{2}}e^{-i\theta_k}\\
        u\cos{\frac{\al_k}{2}}
    \end{pmatrix},\label{eq:psi_S}
\end{align}
where $t_e$ and $t_h$ are the transmission coefficients of the processes where the incident electron is transmitted to an electron-like state and to a hole-like state, respectively, and $u^2=\frac{1}{2}(1+\frac{\sqrt{E^2-\Delta^2}}{E})$ and $v^2=\frac{1}{2}(1-\frac{\sqrt{E^2-\Delta^2}}{E})$. The wavefunctions at $E<\Delta$ can be derived by analytic continuation. 

Due to the facts that the Hamiltonian is first order in momentum and that a delta function potential barrier is present, the wavefunction is in general discontinuous at the interface. To derive the boundary condition, we first derive the expression for the probability current on both sides, which has to be continuous. The probability current in the $x$-direction is given by $j_{x}=\sum\Psi^{\dagger}\hat{v}_{x}\Psi$, where $\hat{v}_{x}=\frac{\partial H}{\partial k_{x}}$ and the sum is over momenta. On both sides, this leads to the expression $j_{x}=v_{F}\sum\Psi^{\dagger}\tau_{z}\otimes\sigma_{x}\Psi$, where $\tau_i$'s are the Pauli matrices acting on Nambu space. Without the delta function potential barrier, the continuity of the probability current simply results in $\Psi_{S}=\Psi_{N}$. To include the effects of the interface barrier, we assume $\Psi_{S}-\Psi_{N}=\tau_{z}\otimes\sigma_{x}\Psi(0)$. From the continuity of $j_x$ at the interface, we find $\Psi(0)=-\frac{i}{2}Z(\Psi_{S}+\Psi_{N})$, where $Z=U_{0}/\hbar v_{F}$.
%\begin{eqnarray}
%\Psi_{S}-\frac{v_{F}^{N}}{v_{F}^{S}}\Psi_{N}=-i\tau_{z}\otimes\sigma_{x}\Psi_{0} &=&-i\tau_{z}\otimes\sigma_{x}\frac{1}{2}Z(\Psi_{S}+\frac{v_{F}^{N}}{v_{F}^{S}}\Psi_{N}),\nonumber\\\\
%\frac{v_{F}^{N}}{v_{F}^{S}}[1-i\tau_{z}\otimes\sigma_{x}\frac{1}{2}Z]\Psi_{N} &=&[1+i\tau_{z}\otimes\sigma_{x}\frac{1}{2}Z]\Psi_{S}
%\end{eqnarray}

\begin{figure}
    \centering
    \includegraphics[width=0.5\linewidth]{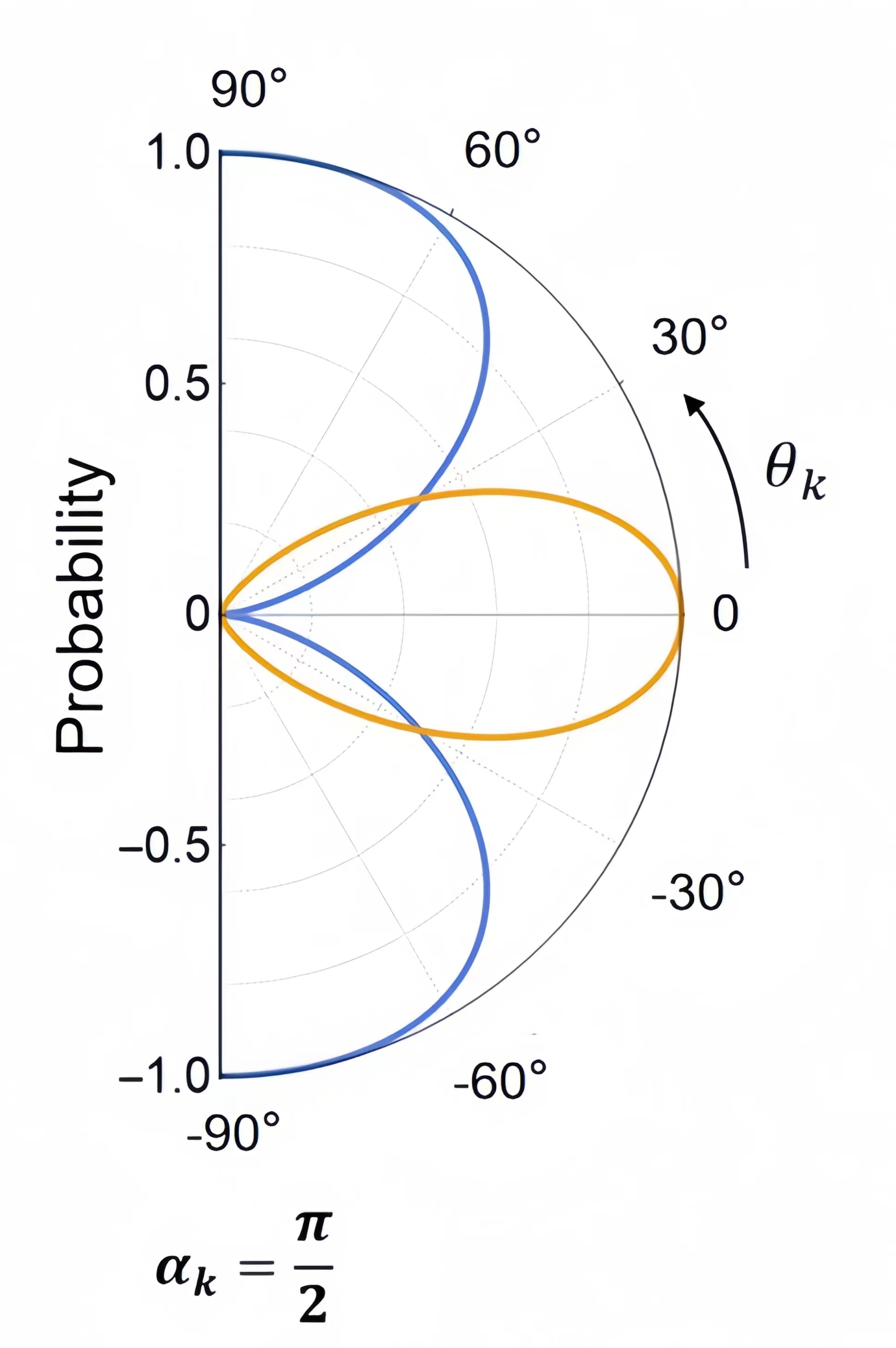}
    \caption{The normal reflection (blue) and Andreev reflection (orange) probabilities in polar coordinates in the $k_z=0$ plane at $E<\Delta$.}
    \label{fig:andreev}
\end{figure}

\begin{figure}[t]
    \centering
    \includegraphics[width=0.98\linewidth]{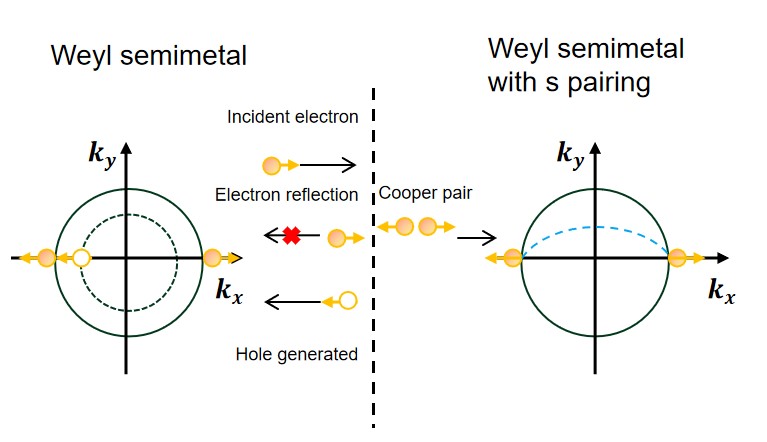}
    \caption{Schematic of how pseudospin-momentum locking leads to perfect Andreev reflection.}
    \label{fig:SW}
\end{figure}

Now we are ready to match the wavefunctions on the N side and S side at the interface. Taking the wavefunctions on both sides into the boundary condition, we get
\begin{align}
    &(1-iZ\tau_z\otimes\sigma_x)\left[\begin{pmatrix}
        \cos{\frac{\alpha_k}{2}}\\
        \sin{\frac{\alpha_k}{2}}e^{i\theta_k}\\
        0\\
        0
    \end{pmatrix}+r_e\begin{pmatrix}
        \cos{\frac{\alpha_k}{2}}\\
        -\sin{\frac{\alpha_k}{2}}e^{-i\theta_k}\\
        0\\
        0
    \end{pmatrix}+r_h\begin{pmatrix}
        0\\
        0\\
        -\sin{\frac{\alpha_k}{2}}e^{i\theta_k}\\
        \cos{\frac{\alpha_k}{2}}
    \end{pmatrix}\right]\nonumber\\
    &=(1+iZ\tau_z\otimes\sigma_x)\left[t_e\begin{pmatrix}
       u\cos{\frac{\al_k}{2}}\\
       u\sin{\frac{\al_k}{2}}e^{i\theta_k}\\
       -v\sin{\frac{\al_k}{2}}e^{i\theta_k}\\
       v\cos{\frac{\al_k}{2}}
    \end{pmatrix}+t_h\begin{pmatrix}
        v\cos{\frac{\al_k}{2}}\\
        -v\sin{\frac{\al_k}{2}}e^{-i\theta_k}\\
        u\sin{\frac{\al_k}{2}}e^{-i\theta_k}\\
        u\cos{\frac{\al_k}{2}}
    \end{pmatrix}\right].
\end{align}
Then we solve the four equations for the four transport coefficients, focusing on $r_e$ and $r_h$. The correctness of the results can be tested by the probability current conservation, i.e. $|r_e|^2+|r_h|^2+\frac{w_e}{v_F}|t_e|^2+\frac{w_h}{v_F}|t_h|^2=1$, where $w_e$ and $w_h$ are the velocities of the electron-like and hole-like states on the S side, respectively. In Fig.\ref{fig:andreev}, we plot the normal reflection probability $|r_e|^2$ and Andreev reflection probability $|r_h|^2$ in polar coordinates in the $k_z=0$ plane at $E<\Delta$. We see that at normal incidence (i.e. $\al_k=\pi/2$ and $\theta_k=0$), the incident electron is completely reflected to a hole state. Such perfect Andreev reflection indicates Klein tunneling\cite{SLee2019}. Klein tunneling in the Weyl junction can be understood as a consequence of pseudospin-momentum locking, as shown in Fig.\ref{fig:SW}. At oblique incidence, there is no Klein tunneling. Actually, in that case, another interesting phenomenon called Imbert-Fedorov shift could occur\cite{Fedorov1955,Imbert1972,Jiang2015,Yang2015,Wang2017,Wang2019}, which is beyond the scope of this paper.

To show Klein tunneling in the heterostructure experimentally, imagine that a beam of Weyl electrons is incident from left, with the beam center incident normally. Assume the beam is Gaussian, with the angular distribution
\bea
f_{\theta_{k}}&=&\frac{1}{\sqrt{\pi}\varrho}e^{-\theta^{2}_{k}/\varrho^{2}},\nonumber\\
f_{\alpha_{k}}&=&\frac{1}{\sqrt{\pi}\varrho}e^{-(\alpha_k-\frac{\pi}{2})^{2}/\varrho^{2}}
\eea
where $f_{\theta_{k}}$ and $f_{\al_{k}}$ are the angular distributions of the beam along $\theta_k$ and $\al_k$ direction, respectively, and $\varrho$ parameterizes the spread of the distribution. In the case $\varrho\ll\pi$, where only $\al_k\sim\pi/2$ and $\theta_{k}\sim0$ channel contributes to the transport, perfect Andreev reflection will be manifested. The differential conductance through the interface is given by
\begin{eqnarray}
&&G=\frac{dI}{dV}\nonumber\\
&&=G_{0}\int_{0}^{\pi}\int^{\frac{\pi}{2}}_{-\frac{\pi}{2}}(1-|r_{e}|^{2}+|r_{h}|^{2}) f_{\theta_{k}}f_{\alpha_k}\cos\theta_{k}\sin{\alpha_k} d\theta_{k}d\alpha_k.
\end{eqnarray}
We show the normalized differential conductance as a function of $E$ for different $\varrho$'s. As one can see, $G$ is approximately twice $G_0$, the differential conductance at $E\gg\Delta$. This manifests Klein tunneling in the Weyl junction.

\begin{figure}[htb]
    \centering
    \includegraphics[width=1\linewidth]{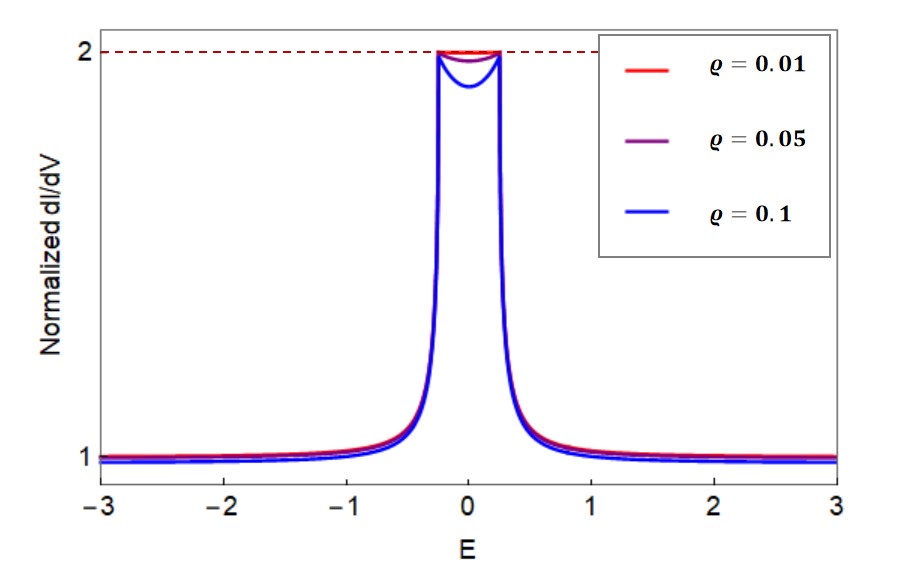}
    \caption{Simulated conductance spectra for different $\varrho$'s (i.e. full width at half maximum of Gaussian angular distribution).}\label{fig:conductance}
\end{figure}

\textit{Anti-Klein tunneling in DWSM-sDWSM junctions.}---Now we study the tunneling in junctions where the single WSMs are replaced by DWSMs, i.e. the heterostructure has a DWSM on the N side and an sDWSM on the S side. The Hamiltonian of the heterostructure resembles Eq.(\ref{eq:H_hetero}), but with $v_F\bk\cdot\boldsymbol{\sigma}$ replaced by
\bea
H_0^d(\bk)=\left[\begin{array}{cc}vk_z&\lambda(k_x-ik_y)^2\\\lambda(k_x+ik_y)^2&-vk_z\end{array}\right].
\eea
In the following, we will assume $\lambda=1$, which does not affect our conclusions. The eigenenergies of $H_0^d(\bk)$ are
\bea
E_{0\pm}(\bk)=\pm\sqrt{v^2k_z^2+(k_x^2+k_y^2)^2}.
\eea

\begin{figure}[t]
    \centering
    \includegraphics[width=\linewidth]{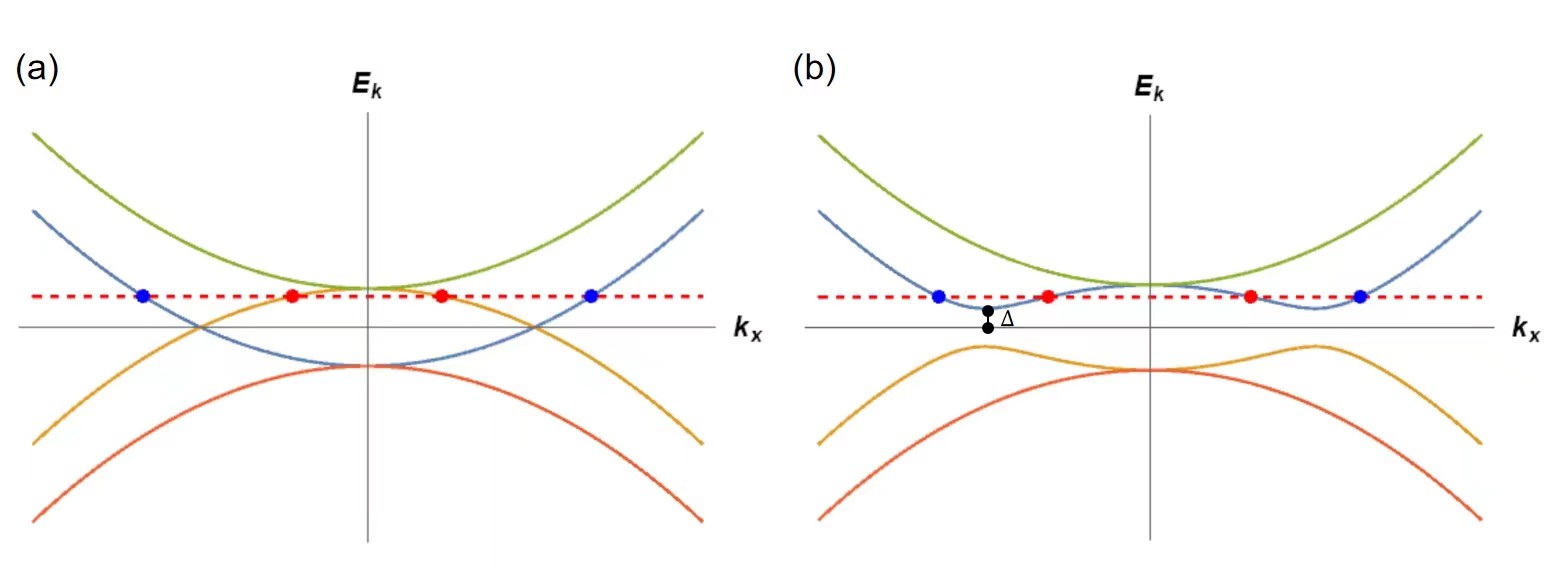}
    \caption{(a) Dispersion of a DWSM along the $k_x$-axis in Nambu space. (b) Dispersion of an sDWSM along the $k_x$-axis. The blue (red) dots indicate the eletron-like (hole-like) propagating wavefunctions at $\Delta<E<\mu$. }
    \label{fig:DW}
\end{figure}

The dispersion (for a fixed $\mu$) in the Nambu space is shown in Fig.\ref{fig:DW}(a). Given a fixed energy $E$ and fixed $k_y$ and $k_z$, for (single) WSMs, there are two possible $k_x$'s. However, for DWSMs, there are four possible $k_x$'s, because the equation $(k_x^2+k_y^2)^2=\rm{const}$ has four solutions. Assume $E=\sqrt{v^2k_z^2+(k_x^2+k_y^2)^2}-\mu>0$, the four solutions of $k_x$ are: $k_x=\pm k_{ix}$ where $k_{ix}=\sqrt{\sqrt{(E+\mu)^2-v^2k_z^2}-k_y^2}$ is a positive number, and $k_x=\pm i\kappa_{x}$ where $
\kappa_{x}=\sqrt{\sqrt{(E+\mu)^2-v^2k_z^2}+k_y^2}=\sqrt{k_{ix}^2+2k_y^2}$ is a positive number. The wavefunctions corresponding to each solution are:
\bea
k_{ix}&:& \left(\begin{array}{c}
                1 \\
                \eta e^{2i\phi_k}
              \end{array}\right)e^{ik_{ix}x};~~~
-k_{ix}: \left(\begin{array}{c}
                1 \\
                \eta e^{-2i\phi_k}
              \end{array}\right)e^{-ik_{ix}x};\nonumber\\
-i\kappa_x&:&\left(\begin{array}{c}
                1 \\
                -h_+
              \end{array}\right)e^{\kappa_x x};~~~~~~~~~~
i\kappa_x:\left(\begin{array}{c}
                1 \\
                -h_-
              \end{array}\right)e^{-\kappa_x x},
\eea
where $\tan\phi_k=k_y/k_{ix}$, $\eta=\frac{E+\mu-vk_z}{k_{ix}^2+k_y^2}=\sqrt{\frac{E+\mu-vk_z}{E+\mu+vk_z}}$, $h_+=\frac{E+\mu-vk_z}{(\kappa_x+k_y)^2}=\frac{(\kappa_x-k_y)^2}{E+\mu+vk_z}$, and $h_-=\frac{E+\mu-vk_z}{(\kappa_x-k_y)^2}$. For the hole Hamiltonian on the N side, the eigenenergies are $E=\pm\sqrt{v^2k_z^2+(k_x^2+k_y^2)^2}+\mu$. For $0<E<\mu$, we take the solution $E=\mu-\sqrt{v^2k_z^2+(k_x^2+k_y^2)^2}$. There are also four solutions of $k_x$ for a given $E$, $k_y$ and $k_z$: $k_x=\pm k_h$ where $k_h=\sqrt{\sqrt{(\mu-E)^2-v^2k_z^2}-k_y^2}$ is a positive number, and $k_x=\pm i\kappa_{h}$ where $\kappa_{h}=\sqrt{\sqrt{(\mu-E)^2-v^2k_z^2}+k_y^2}=\sqrt{k_{h}^2+2k_y^2}$ is a positive number. The wavefunctions corresponding to each solution are:
\bea
k_{h}&:& \left(\begin{array}{c}
                -\eta' e^{2i\phi_h} \\
                1
              \end{array}\right)e^{ik_{h}x};~~
-k_{h}: \left(\begin{array}{c}
                -\eta' e^{-2i\phi_h} \\
                1
              \end{array}\right)e^{-ik_{h}x};\nonumber\\
-i\kappa_h&:&\left(\begin{array}{c}
                h'_+ \\
                1
              \end{array}\right)e^{\kappa_h x};~~~~~~~~~~~~~~~
i\kappa_h:\left(\begin{array}{c}
                h'_-\\
                1
              \end{array}\right)e^{-\kappa_h x},
\eea
where $\tan\phi_h=k_y/k_h$, $\eta'=\sqrt{\frac{-E+\mu-vk_z}{-E+\mu+vk_z}}$, $h'_+=\frac{-E+\mu-vk_z}{(\kappa_h+k_y)^2}$ and $h'_-=\frac{-E+\mu-vk_z}{(\kappa_h-k_y)^2}$.

Therefore, on the N side, the wavefunction is
\bea
&&\psi_N=\left(\begin{array}{c}
               1 \\
               \eta e^{2i\phi_k} \\
               0 \\
               0
             \end{array}\right)e^{ik_{ix}x}+
        r_e\left(\begin{array}{c}
               1 \\
               \eta e^{-2i\phi_k} \\
               0 \\
               0
             \end{array}\right)e^{-ik_{ix}x}\nonumber\\
        &&+r_h\left(\begin{array}{c}
               0 \\
               0 \\
               -\eta' e^{2i\phi_h} \\
               1
             \end{array}\right)e^{ik_{h}x}
        +c_e\left(\begin{array}{c}
               1 \\
               -h_+ \\
               0 \\
               0
             \end{array}\right)e^{\kappa_x x}+
        c_h\left(\begin{array}{c}
               0 \\
               0 \\
               h'_+ \\
               1
             \end{array}\right)e^{\kappa_h x}
\eea
On the S side, the eigenenergies are
\bea
E(\bk)=\pm\sqrt{(E_{0+}(\bk)\pm\mu)^2+\Delta^2},
\eea
as shown in Fig.\ref{fig:DW}(b), and the wavefunction is
\bea
\psi_S&&=t_e\left(\begin{array}{c}
                  u \\
                  u\eta e^{2i\phi_k} \\
                  -v\eta e^{2i\phi_k} \\
                  v
                \end{array}\right)e^{iq_ex}+
        t_h\left(\begin{array}{c}
                  v \\
                  v\eta e^{-2i\phi_h} \\
                  -u\eta e^{-2i\phi_h} \\
                  u
                \end{array}\right)e^{-iq_h x}\nonumber\\
        &&+d_e\left(\begin{array}{c}
               u \\
               -uh_- \\
               vh_- \\
               v
             \end{array}\right)e^{-\kappa' x}+
        d_h\left(\begin{array}{c}
               v \\
               -vh'_- \\
               uh'_- \\
               u
             \end{array}\right)e^{-\kappa'_h x},
\eea
where $q_e=\sqrt{\mu+\sqrt{E^2-\Delta^2}}$, $q_h=\sqrt{\mu-\sqrt{E^2-\Delta^2}}$, $u^2=\frac{1}{2}(1+\frac{\sqrt{E^2-\Delta^2}}{E})$ and $v^2=\frac{1}{2}(1-\frac{\sqrt{E^2-\Delta^2}}{E})$. Note that unlike Eq.(\ref{eq:psi_S}) in the single WSM case, there are four possible solutions on the S side, including two evanescent waves. The boundary conditions are: (1) the wavefunction is continuous, i.e. $\psi_N(0)=\psi_S(0)$; and (2) the derivative of the wavefunctions satisfies the condition from integrating the eigenequations from $0^-$ to $0^+$:
\bea
\psi_{Si}'(0)-\psi'_{Ni}(0)&=&\Lambda\psi_{Nj}(0),
\eea
where for $i=1$, 2, 3, and 4, $j$ takes the values 2, 1, 4, and 3, respectively.
Now we look at the case of normal incidence, where $k_y=k_z=0$. We seek for supra-gap solution, i.e. $\Delta<E<\mu$. In this case, $\phi_k=\phi_h=0$, $\eta=\eta'=1$, $h_+=h_-=1$, $k_{ix}=\sqrt{E+\mu}$, $\kappa_x=\sqrt{E+\mu}$, $k_h=\sqrt{\mu-E}$, $\kappa_h=\sqrt{\mu-E}$, $h'_+=h'_-=1$, $\kappa'=q_e$, and $\kappa'_h=q_h$. Then there are four equations from $\psi_N(0)=\psi_S(0)$, and another four equations are from the equations for the derivatives.

\begin{figure}[t]
    \centering
    \includegraphics[width=0.98\linewidth]{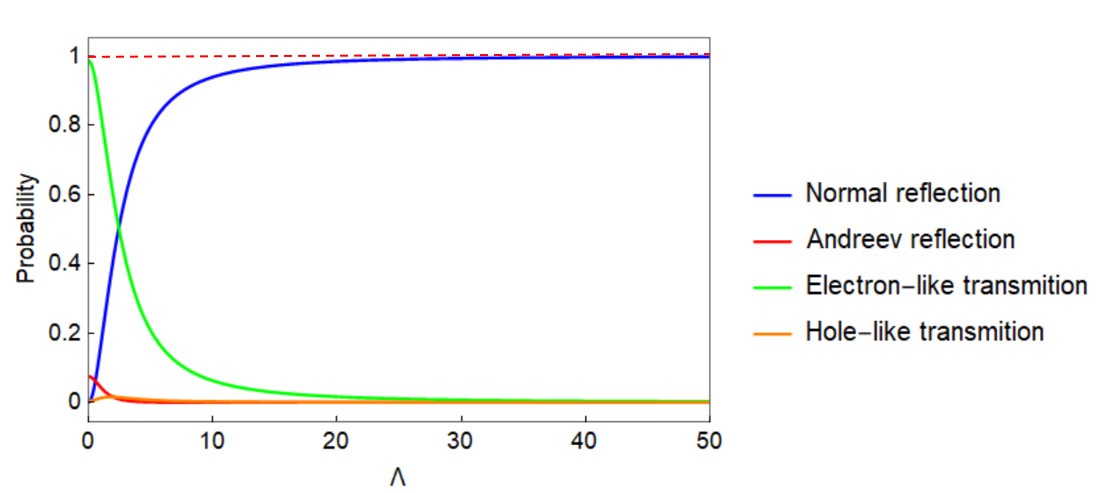}
    \caption{Reflection and transmission probabilities for the normal incidence in a DWSM/sDWSM junction.}
    \label{fig:double-probability}
\end{figure}

Solving the eight equations for the eight coefficients $r_e$, $r_h$, $t_e$, $t_h$, $c_e$, $c_h$, $d_e$ and $d_h$, the first four of which are reflection and transmission coefficients while the last four are coefficients of evanescent waves, we show the reflection and transmission probabilities in Fig.\ref{fig:double-probability}. As can be seen, at large $\Lambda$, incident electrons are completely reflected as electrons and no tunneling occur at all. This phenomenon is called the anti-Klein tunneling, which has been predicted and observed in bilayer graphene\cite{NatureKatsnelson,Elahi2024}.

\textit{Discussion.}---We have proposed a new platform -- WSM-based superconducting junctions -- to observe Klein and anti-Klein tunneling. One can envision that other similar systems may show similar behavior. For example, pseudospin-1 semimetal-based NS junctions may exhibit all-angle-perfect-tunneling, as pseudospin-1 fermions themselves can show when their energy is half the potential barrier height\cite{Shen2010,Urban2011,Xu2014,Fang2016}. If the WSMs in our setup are replaced by triple WSMs, Klein tunneling could also occur. These studies are left for future work.

{\it Acknowledgments.}---This work was supported by National Key R\&D Program of China (Grant No. 2022YFA1402802, No. 2018YFA0306001), National Natural Science Foundation of China (Grant No. 12004442, No. 92165204), Guangdong Basic and Applied Basic Research Foundation (Grant No. 2021B1515130007), Shenzhen Natural Science Fund (the Stable Support Plan Program 20220810130956001), Shenzhen International Quantum Academy, and Guangdong Provincial Key Laboratory of Magnetoelectric Physics and Devices (No. 2022B1212010008).
\bibliography{Weyl_Klein_ref}
\bibliographystyle{apsrev4-2}

\end{document}